\documentclass[prb,aps,showpacs,twocolumn,floatfix]{revtex4-1}
\usepackage{amsmath,amsfonts,amssymb,bm}
\newcommand{\bS}{{\bm S}}
\newcommand{\bk}{{\bm k}}
\newcommand{\bp}{{\bm p}}
\newcommand{\bsig}{{\bm\sigma}}
\newcommand{\bG}{{\bm\Gamma}}
\newcommand{\e}{{\rm e}}
\newcommand{\ii}{{\rm i}}
\newcommand{\cd}{c^\dag}
\newcommand{\dd}{d^\dag}
\newcommand{\ua}{\uparrow}
\newcommand{\da}{\downarrow}
\newcommand{\vac}{\vert0\rangle}
\newcommand{\FS}{\vert{\rm FS}\rangle}
\newcommand{\hFS}{\langle{\rm FS}\vert}

\newcommand{\ve}{\varepsilon}
\begin{document}
\title{Kondo effect in the presence of spin-orbit coupling}
\author{L. Isaev$^1$}
\author{D. F. Agterberg$^2$}
\author{I. Vekhter$^1$}
\affiliation{$^1$Department of Physics and Astronomy,
                 Louisiana State University, Baton Rouge LA 80703 \\
             $^2$Department of Physics, University of Wisconsin,
                 Milwaukee WI 53201}
\begin{abstract}
 We study the $T=0$ Kondo physics of a spin-$1/2$ impurity in a
 non-centrosymmetric metal with spin-orbit interaction. Within a simple
 variational approach we compute ground state properties of the system for an
 {\it arbitrary} form of spin-orbit coupling consistent with the crystal
 symmetry. This coupling produces an unscreened impurity magnetic moment and
 can lead to a significant change of the Kondo energy. We discuss implications
 of this finding both for dilute impurities and for heavy-fermion materials
 without inversion symmetry.
\end{abstract}
\maketitle

\paragraph*{Introduction.--}
Kondo effect, i.e. screening of the impurity magnetic moment by the Fermi
sea of itinerant electrons, is one of the best-known examples of
correlations-driven phenomena in condensed matter physics \cite{mahan-1990}. A
system involving a periodic array of such impurities interacting with
conduction electrons (the so-called Kondo lattice) is believed to provide a
minimal model for heavy-fermion compounds \cite{hewson-1993}. Historically the
Kondo screening was detected via resistivity measurements in dilute magnetic
alloys, but recent advances in scanning tunneling spectroscopy allowed
observation of this phenomenon on the atomic scale
\cite{madhavan-1998,li-1998,pruser-2011} and manipulation of individual Kondo
resonances \cite{tsukahara-2011}.

Details of the band structure of the host metal usually do not qualitatively
influence the Kondo ground state, although they affect characteristic energy
scales of the problem such as the Kondo temperature, $T_K$, below which the
impurity spin is screened. Similarly, in the presence of spin-orbit scattering
when spin is not a good quantum number, classification of the states by parity
still allows mapping of the impurity problem onto a Kondo model with
essentially same parameters but without the local spin-orbit interaction (SOI)
\cite{meir-1994}, in agreement with experiment \cite{bergmann-1986,wei-1989}.

In non-centrosymmetric materials a distinct non-local (dependent on the
gradients of the crystal potential) SOI appears. This interaction is odd in
electron momentum and couples it to the electron spin \cite{bir-1974}. The
influence of this type of SOI on manifestations of the Kondo effect was
discussed only recently \cite{malecki-2007,zarea-2011,zitko-2011,feng-2011} in
quasi two-dimensional (2D) systems for specific cases of Rashba or Dresselhaus
SOI, and in the context of topological insulators \cite{feng-2010,zitko-2010}.

In Ref. \onlinecite{malecki-2007} it was concluded that, to lowest order, the
Rashba SOI only leads to a rescaling of the electron bandwidth and leaves the
Kondo temperature essentially unchanged. A similar verdict was reached in Ref.
\onlinecite{zarea-2011} in the framework of the Anderson model for a
half-filled $f$-band. However these results rely heavily on the specific form
of the Rashba SOI term and 2D single-particle density of states. This
particular combination allows reduction of the Kondo Hamiltonian with SOI
(equivalent to a multichannel problem, see below) to a single-channel model
without spin-orbit coupling. What happens with Kondo screening in more
realistic and interesting cases, e.g. three-dimensional materials without
inversion symmetry or systems with a non-Rashba SOI that do not allow the above
simplification has not been explored.

In this Communication we consider a single spin-$1/2$ impurity interacting with
a system of electrons in a non-centrosymmetric metal at zero temperature. Due
to the explicit inversion symmetry breaking, the single-particle Hamiltonian
that describes the conduction band contains an odd in momentum spin-orbit term
compatible with the crystal symmetry \cite{samokhin-2009}. We determine the
ground state properties of the resulting Kondo Hamiltonian by generalizing
Yosida's variational method \cite{yosida-1966} to take into account the
spin-orbit splitting of the Fermi surface (FS), as well as all values of the
total spin of the electrons and the impurity. In contrast with previous works
\cite{malecki-2007,zarea-2011,zitko-2011,feng-2011} our analysis is valid for
any form of SOI and the electron band structure, and incorporates the
essentially multichannel nature of the problem. We give general expressions for
the Kondo binding energy and show that the SOI may lead to an enhancement of
the Kondo effect compared to that of a centrosymmetric material with the same
parameters. Because the SOI breaks $SU(2)$ symmetry, the impurity spin no
longer forms a singlet with the Fermi sea and is only partially screened. This
conclusion is qualitatively similar to the situation in 2D helical metals
\cite{feng-2010}. Since our goal is to investigate only effects associated with
SOI we ignore possible spin anisotropy terms analogous to those appearing in
the study of impurities near sample surfaces \cite{usaghy-2007}.

We first set up the variational framework, and then present results for the
Kondo binding energy, total spin in the ground state, and the impurity spin
susceptibility.

\paragraph*{Variational formalism.--}
The Kondo model describes a localized magnetic impurity interacting with a
single band of conduction electrons
\begin{displaymath}
 H=\sum_\bk\ve_{\alpha\beta}(\bk)\cd_{\bk\alpha}c_{\bk\beta}+
 J_K\bS\bsig_{\alpha\beta}\cd_{i_0\alpha}c_{i_0\beta}.
\end{displaymath}
This Hamiltonian is defined on a lattice with $N$ sites; $\bS$ is the impurity
spin ($S=1/2$) located at site $i_0$, $\bsig_{\alpha\beta}$ are Pauli matrices,
$\cd_{i\alpha}$ creates a fermion at site $i$ with spin $\alpha=(\ua,\da)$
($\cd_{\bk\alpha}=\sqrt{1/N}\sum_i\e^{-\ii\bk{\bm x}_i}\cd_{i\alpha}$ is its
momentum space counterpart), and $\ve_{\alpha\beta}(\bk)$ is the
single-electron dispersion. We take $J_K>0$, assume summation over repeated
indices, and set $\hbar\equiv1$.

For a single band with SOI the matrix $\ve_{\alpha\beta}(\bk)$ can be written
\cite{samokhin-2009} as:
\begin{displaymath}
 \ve_{\alpha\beta}(\bk)=\epsilon_\bk\delta_{\alpha\beta}+\bG_\bk
 \bsig_{\alpha\beta}.
\end{displaymath}
The scalar $\epsilon_\bk$ is the dispersion without SOI. The latter enters
through the real pseudovector $\bG_\bk=-\bG_{-\bk}$ which is determined by the
point group symmetry of the crystal. It is convenient to diagonalize
$\ve_{\alpha\beta}$ explicitly by introducing the helicity basis
$c_{\bk\alpha}=(U_\bk)_{\alpha\lambda}d_{\bk\lambda}$ with $\lambda=\pm1$ and
unitary matrix $U_\bk$ such that
$\bG_\bk U_\bk^\dag\bsig U_\bk=\sigma^z\vert\bG_\bk\vert$. In this
representation the band energy is diagonal,
$\ve_{\bk\lambda}=(U_\bk^\dag)_{\lambda\alpha}\ve_{\alpha\beta}(\bk)
(U_\bk)_{\beta\lambda}=\epsilon_\bk+\lambda\vert\bG_\bk\vert$. Note that
$\bG_\bk$ breaks parity but preserves time-reversal, hence
$\ve_{\bk\lambda}=\ve_{-\bk,\lambda}$ because of the Kramers theorem. Now we
can rewrite the Kondo Hamiltonian as:
\begin{align}
 H=&\sum_{\bk,\lambda}\ve_{\bk\lambda}\dd_{\bk\lambda}d_{\bk\lambda}+
 \label{kondo_hamiltonian} \\
 &+\frac{J_K}{N}\sum_{\bk^\prime,\bk}\bS\bigl(U^\dag_{\bk^\prime}\bsig
 U_\bk\bigr)_{\lambda^\prime\lambda}\dd_{\bk^\prime\lambda^\prime}
 d_{\bk\lambda}. \nonumber
\end{align}

To understand the influence of SOI on the Kondo screening, we use the
Yosida-like \cite{yosida-1966} trial wavefunction
\begin{equation}
 \vert\psi\rangle=\sum_{\bk,M,\lambda}A_{\bk M\lambda}
 \theta(\ve_{\bk\lambda}-\ve_F)\vert M\rangle\dd_{\bk\lambda}\FS,
 \label{trial_state}
\end{equation}
where $A_{\bk M\lambda}$ are variational amplitudes, $M=(\ua,\da)$ labels
impurity states and $\FS$ is the filled Fermi sea
\begin{displaymath}
 \FS=\prod_{\ve_{\bk+}<\ve_F}\dd_{\bk+}\prod_{\ve_{\bk-}<\ve_F}\dd_{\bk-}\vac.
\end{displaymath}
The Heaviside function $\theta(\ve_{\bk\lambda}-\ve_F)$ limits summation to
the energies above the Fermi level. The expectation value of the Hamiltonian
Eq. \eqref{kondo_hamiltonian} in the state $\vert\psi\rangle$ of Eq.
\eqref{trial_state} is
\begin{align}
 \langle\psi\vert H\vert\psi\rangle=&\sum
 A^*_{\bk^\prime M^\prime\lambda^\prime}A_{\bk M\lambda}
 \theta(\ve_{\bk^\prime\lambda^\prime}-\ve_F)\theta(\ve_{\bk\lambda}-\ve_F)
 \times \nonumber \\
 \times&\biggl[\ve_{\bk\lambda}
 \delta_{\lambda^\prime\lambda}\delta_{\bk^\prime\bk}\delta_{M^\prime M}+
 \frac{J_K}{N}\bS_{M^\prime M}\bigl(U^\dag_{\bk^\prime}\bsig U_\bk
 \bigr)_{\lambda^\prime\lambda}\biggr], \nonumber
\end{align}
with the implicit summation over all indices in the r.h.s. In this expression
we omitted the $A$-independent ground state energy of the Fermi sea,
$E_0=\hFS H\FS$. Computing the expectation value of the Kondo interaction
requires decoupling of the product
\begin{align}
 \hFS d_{\bp^\prime\lambda^\prime}\dd_{\bk^\prime\alpha}&d_{\bk\beta}
 \dd_{\bp\lambda}\FS=\theta(\ve_{\bp^\prime\lambda^\prime}-\ve_F)\times
 \nonumber \\
 &\times\bigl[\delta_{\alpha\beta}\delta_{\bk\bk^\prime}\delta_{\bp\bp^\prime}
 \delta_{\lambda\lambda^\prime}\theta(\ve_F-\ve_{\bk\alpha})+ \nonumber \\
 &\qquad+\delta_{\bk\bp}\delta_{\bk^\prime\bp^\prime}\delta_{\beta\lambda}
 \delta_{\alpha\lambda^\prime}\theta(\ve_{\bp\lambda}-\ve_F)\bigr]. \nonumber
\end{align}
In this equation the first term has the form
$\sum_{\bk,\alpha}\theta(\ve_F-\ve_{\bk\alpha})\bigl(U^\dag_\bk\bsig
U_\bk\bigr)_{\alpha\alpha}=2\hFS\bS_e\FS$, where
$\bS_e=(1/2)\sum_\bk\bsig_{\alpha\beta}\cd_{\bk\alpha}c_{\bk\beta}$ is
the total electron spin; $\hFS\bS_e\FS=0$ due to the time-reversal symmetry.

Minimizing $\langle\psi\vert H\vert\psi\rangle$ w.r.t. $A_{\bk M\lambda}$, one
obtains an eigenvalue equation:
\begin{align}
 (\ve_{k\lambda}-E)&A_{\bk M\lambda}\theta(\ve_{\bk\lambda}-\ve_F)=
 -J_K\bS_{MM^\prime}\theta(\ve_{\bk\lambda}-\ve_F)\times \nonumber \\
 &\times\frac{1}{N}\sum_{\bp,\eta}\theta(\ve_{\bp\eta}-\ve_F)
 \bigl(U^\dag_\bk\bsig U_\bp\bigr)_{\lambda\eta}A_{\bp M^\prime\eta},
 \label{main_eq}
\end{align}
where again the summation over doubly repeated indices is assumed. To proceed
further, we introduce
\begin{displaymath}
 B_{\bk M\lambda}=\sum_\eta\bigl(U_\bk\bigr)_{\lambda\eta}
 \theta(\ve_{\bk\eta}-\ve_F)A_{\bk M\eta},
\end{displaymath}
which allows us to rewrite Eq. \eqref{main_eq} in the form
\begin{align}
 &B_{\bk M\alpha}=-J_K\bS_{MR}\times \label{wavefunction} \\
 &\times\biggl[\sum_\lambda\bigl(
 U_\bk\bigr)_{\alpha\lambda}\frac{\theta(\ve_{\bk\lambda}-\ve_F)}
 {\ve_{\bk\lambda}-E}\bigl(U^\dag_\bk\bigr)_{\lambda\beta}\biggr]
 \bsig_{\beta\gamma}\biggl(\frac{1}{N}\sum_\bp B_{\bp R\gamma}\biggr).
 \nonumber
\end{align}
This object plays the role of the ground-state wavefunction for the system.
Due to the $\theta$-function in the definition of $B_{\bk M\alpha}$ all
$\bk$-summations are over the entire Brillouin zone.

We shall now use Eqs. \eqref{trial_state}, \eqref{main_eq},
\eqref{wavefunction} to compute the Kondo energy, total spin of the system and
impurity magnetic susceptibility in the most general form. Then we apply
obtained expressions to several instructive examples: (i) quasi-2D systems with
symmetry $C_{4v}$ (with Rashba or Dresselhaus SOI), and (ii) cubic crystals
with symmetry $T$ or $O$.

\paragraph*{Kondo energy.--}
The energy eigenvalue $E$ in Eq. \eqref{main_eq} is obtained by summing Eq.
\eqref{wavefunction} over $\bk$
\begin{align}
 &X_{M\alpha}= \nonumber \\
 &=-\frac{J_K}{N}\bS_{MR}\sum_{\bk,\lambda}\biggl[\bigl(
 U_\bk\bigr)_{\alpha\lambda}\frac{\theta(\ve_{\bk\lambda}-\ve_F)}
 {\ve_{\bk\lambda}-E}\bigl(U^\dag_\bk\bigr)_{\lambda\beta}\biggr]
 \bsig_{\beta\gamma}X_{R\gamma}, \nonumber
\end{align}
with $X_{M\alpha}=(1/N)\sum_\bk B_{\bk M\alpha}$. The $\lambda$-dependent
terms between two $U$-matrices can be decomposed as
\begin{displaymath}
 \theta(\ve_{\bk\lambda}-\ve_F)\delta_{\lambda^\prime\lambda}/
 (\ve_{\bk\lambda}-E)=\delta_{\lambda^\prime\lambda}\kappa_++
 \sigma^z_{\lambda^\prime\lambda}\kappa_-,
\end{displaymath}
where
\begin{equation}
 \kappa_\pm(\bk)=\frac{1}{2}\biggl(
 \frac{\theta(\ve_{\bk+}-\ve_F)}{\ve_{\bk+}-E}\pm
 \frac{\theta(\ve_{\bk-}-\ve_F)}{\ve_{\bk-}-E}\biggr).
 \label{kappas}
\end{equation}
Because $U_\bk\sigma^zU^\dag_\bk=\bG_\bk\bsig/\vert\bG_\bk\vert$ and $\bG_\bk$
is odd, while $\ve_{\bk\lambda}(\bk)$ is even in $\bk$, the term containing
$\sigma^z\kappa_-$ does not contribute to the sum and we find
\begin{equation}
 X_{M\alpha}=-\frac{J_K}{N}\bS_{MR}\bsig_{\alpha\beta}\sum_\bk\kappa_+
 X_{R\beta}.
 \label{yosida_eq}
\end{equation}
Clearly, the lowest-energy solution has the ``singlet'' structure in the
helicity space: $X_{M\alpha}=\bigl(\delta_{M\ua}\delta_{\alpha-}-\delta_{M\da}
\delta_{\alpha+}\bigr)/\sqrt{2}$. Then the sum is computed as
\begin{displaymath}
 \frac{1}{N}\sum_\bk\kappa_+=\frac{1}{2}\sum_\lambda
 \int_{\varepsilon_F}^{\varepsilon_F+W}\frac{d\epsilon\,g_\lambda(\epsilon)}
 {\epsilon-E}\approx\frac{g_F^++g_F^-}{2}\ln\frac{W}{\delta E},
\end{displaymath}
where $W$ and $\varepsilon_F\sim W$ are the half-bandwidth and Fermi energy
respectively, $\delta E=\varepsilon_F-E\ll W$, and $g_F^{\lambda}$ is the
density of states (DOS) in the $\lambda$-branch at the Fermi level. From this
expression we finally obtain the energy of the Kondo bound state
\begin{equation}
 \delta E=W\e^{-4/3J_K(g_F^++g_F^-)}.
 \label{kondo_energy}
\end{equation}

When the SOI is absent $g_F^+=g_F^-=g_F^{(0)}$ and Eq. \eqref{kondo_energy}
reduces to the well-known result for the usual Kondo effect \cite{mahan-1990}:
$\delta E^{(0)}=W\exp[-2/3J_K g_F^{(0)}]$. If the
characteristic SOI energy for electrons near the FS is $\Lambda_{SO}\ll\ve_F$,
we expand the DOS at the Fermi level up to the second order
$g_F^\lambda\simeq g_F^{(0)}+\Lambda_{SO}\partial
g^\lambda_F/\partial\Lambda_{SO}+\Lambda_{SO}^2\partial^2g^\lambda_F/2
\partial\Lambda_{SO}^2$, where the derivatives are evaluated at
$\Lambda_{SO}=0$. We estimate in a metal
$\partial g^\lambda_F/\partial\Lambda_{SO}\sim-\lambda g_F^{(0)}/\ve_F$, and
$\partial^2g^\lambda_F/\partial\Lambda_{SO}^2\sim\pm g_F^{(0)}/\ve_F^2$ with
the sign depending on the curvature of the DOS around the FS. Therefore
$(g_F^++g_F^-)/2\sim g_F^{(0)}(1\pm\Lambda^2_{SO}/\ve_F^2)$. While in typical
materials $\Lambda_{SO}/\varepsilon_F\sim 0.1$ and the above correction is only
$\sim1\%$, the exponential form of the Kondo energy, Eq. \eqref{kondo_energy},
makes the effect non-negligible
\begin{equation}
 \delta E/\delta E^{(0)}=\e^{\pm\Lambda_{SO}^2/\ve_F^2J_Kg_F^{(0)}}
 \simeq \e^{\pm\Lambda_{SO}^2/\ve_F J_K} .
 \label{kondo_energy_small_SO}
\end{equation}
Assuming $\Lambda_{SO}\sim J_K$ this gives $\sim 10\%$ change in the Kondo
energy relative to its value $\delta E^{(0)}$ without the SOI.

It is instructive to apply the general expressions \eqref{kondo_energy} and
\eqref{kondo_energy_small_SO} to two examples with parabolic bands (with an
effective mass $m$) and a linear in $\bk$ SOI: (i) quasi-2D tetragonal systems
characterized by Rashba (Dresselhaus) spin-orbit coupling with
$\bG_\bk=\Delta_{SO}[\bk\times{\bm e}_z]$  ($\bG_\bk=\Delta_{SO}\bk$) and
tetragonal axis pointing in the $z$ direction, and (ii) non-centrosymmetric
cubic crystals \cite{samokhin-2009} with $\bG_\bk=\Delta_{SO}\bk$. The coupling
constant $\Delta_{SO}$, which has units of velocity, introduces a natural
energy scale $\epsilon_{SO}=m\Delta^2_{SO}/2$, and is related to $\Lambda_{SO}$
via $\Delta_{SO}\sim\Lambda_{SO}/k_F$ where $k_F$ is the helicity-averaged
Fermi momentum. Consequently, $\epsilon_{SO}\sim\Lambda^2_{SO}/\varepsilon_F$.

In case (i) the DOS per helicity $\lambda$ and for positive energies is given
by $g_\lambda(\epsilon>0)=g_F^{(0)}[1-\lambda\sqrt{\epsilon_{SO}/
(\epsilon_{SO}+\epsilon)}]$ with $g_F^{(0)}=m/2\pi$. Consequently
$g_F^+ +g_F^-=2g_F^{(0)}$ and Eq. \eqref{kondo_energy} yields no correction to
the Kondo energy \cite{malecki-2007,zarea-2011}: $\delta E=\delta E^{(0)}$.
This conclusion is specific {\it solely} to 2D systems with parabolic bands and
linear SOI. Of course, cubic in momentum SOI terms will introduce corrections
of the form \eqref{kondo_energy_small_SO}. In contrast, for case (ii) we have
\begin{displaymath}
 g_\lambda(\epsilon>0)=\frac{m^2\Delta_{SO}}{\pi^2}\biggl(
 \frac{1+\epsilon/2\epsilon_{SO}}{\sqrt{1+\epsilon/\epsilon_{SO}}}-\lambda
 \biggr).
\end{displaymath}
When $\epsilon_{SO}\ll\ve_F$ this DOS leads to an enhancement of the Kondo
energy $\delta E/\delta E^{(0)}=\e^{\epsilon_{SO}/\varepsilon_F J_Kg_F^{(0)}}$,
in agreement with Eq. \eqref{kondo_energy_small_SO}.

It is important to emphasize that Eqs. \eqref{kondo_energy} and
\eqref{kondo_energy_small_SO} correspond to a generally infinite channel Kondo
problem even in the parabolic band approximation. Indeed, without SOI the Kondo
Hamiltonian \eqref{kondo_hamiltonian}
can be reduced to a one-dimensional form which simply reflects the fact that
only electrons with zero orbital angular momentum couple to the impurity
\cite{wiegmann-1981}. When the SOI is taken into account, such reduction is not
always possible because of the $U_\bk$-matrices in Eq.
\eqref{kondo_hamiltonian} which entangle different orbital harmonics. While in
systems with Rashba SOI one can still decouple orbital channels by introducing
suitable linear combinations of $c$-operators and show that only one of them
enters the Kondo term \cite{malecki-2007}, other forms of SOI, e.g. case (ii)
considered above, do not allow such simplification. Thus the validity of Eq.
\eqref{kondo_energy} is only restricted by the variational Ansatz
\eqref{trial_state}.

\paragraph*{Total spin in the ground state.--}
In the standard Kondo problem \cite{mahan-1990} at zero temperature the
impurity is fully screened by the Fermi sea and the net spin of the system
vanishes. This is not the case in the presence of a SOI. Because of the latter,
even without the impurity the electron system has a non-zero spin,
$\hFS\bS_e^2\FS=(1/4)\sum\vert(U^\dag_\bk\bsig
U_\bk)_{\mu\nu}\vert^2\theta(\ve_F-\ve_{\bk\mu})\theta(\ve_{\bk\nu}-\ve_F)$.
This expression is finite due to the mismatch between Fermi surfaces for
different helicities. Therefore, our goal in this part is to compute the
difference between net spins in the Kondo and normal metal phases,
$\langle\psi\vert(\bS+\bS_e)^2\vert\psi\rangle/\langle\psi\vert\psi\rangle-
\hFS\bS_e^2\FS$. Note that due to time-reversal symmetry of the problem, the
total spin polarization along any direction still vanishes.

Using Eq. \eqref{kappas} and the singlet structure of $X_{M\alpha}$ [see
discussion after Eq. \eqref{yosida_eq}], we can rewrite Eq.
\eqref{wavefunction} as
\begin{displaymath}
 B_{\bk M\alpha}=\frac{3}{2}J_K\kappa_+X_{M\alpha}-J_K\kappa_-\bS_{MR}\bigl[
 U_\bk\sigma^zU_\bk^\dag\bsig\bigr]_{\alpha\gamma}X_{R\gamma},
\end{displaymath}
so that the norm of the state \eqref{trial_state} becomes
\begin{align}
 \langle\psi\vert\psi\rangle=&\sum_{\bk M\alpha}\bigl \vert A_{\bk M\alpha}
 \bigr \vert^2\theta(\ve_{\bk\lambda}-\ve_F)=\sum_{\bk M\alpha}\bigl \vert
 B_{\bk M\alpha}\bigr \vert^2= \nonumber \\
 &=\biggl(\frac{3J_K}{2}\biggr)^2\sum_\bk\bigl(\kappa^2_++\kappa^2_-\bigr).
 \nonumber
\end{align}
The cross-terms $\sim\kappa_+\kappa_-$ vanish due to the same argument as that
used in deriving Eq. \eqref{yosida_eq}. Next, we consider the expectation value
of $\bS\bS_e$
\begin{align}
 \langle\psi\vert&\bS\bS_e\vert\psi\rangle=\frac{1}{2}\sum_\bk
 B^*_{\bk M^\prime\alpha^\prime}\bS_{M^\prime M}\bsig_{\alpha^\prime\alpha}
 B_{\bk M\alpha}\equiv \nonumber \\
 &\equiv-\frac{3}{4}\biggl(\frac{3J_K}{2}\biggr)^2\sum_\bk\kappa^2_++
 \frac{J_K^2}{2}\sum_\bk\kappa^2_-X^*_{R^\prime\gamma^\prime}
 T^{R^\prime\gamma^\prime}_{R\gamma}X_{R\gamma}, \nonumber
\end{align}
where again there are no cross-terms and $T^{R^\prime\gamma^\prime}_{R\gamma}=
(S^iS^lS^j)_{R^\prime R}(\sigma^iU_\bk\sigma^zU^\dag_\bk\sigma^lU_\bk\sigma^z
U^\dag_\bk\sigma^j)_{\gamma^\prime\gamma}$. Since $\bS$ is a spin-1/2 operator,
we can evaluate $T$ using the relations
$X^*_{R^\prime\gamma^\prime}\sigma^a_{R^\prime R}\sigma^b_{\gamma^\prime\gamma}
X_{R\gamma}=-\delta_{ab}$ and $\sigma^i\sigma^l\sigma^j=i\varepsilon_{ilj}+
(\delta_{il}\delta_{sj}+\delta_{jl}\delta_{is}-\delta_{ls}\delta_{ij})\sigma^s$
where $\varepsilon_{ilj}$ is the fully antisymmetric tensor:
\begin{displaymath}
 T^{R^\prime\gamma^\prime}_{R\gamma}=\frac{1}{8}\biggl[(2\delta_{R^\prime R}
 \delta_{\gamma^\prime\gamma}-3\bsig_{R^\prime R}\bsig_{\gamma^\prime\gamma})+2
 \sigma^a_{R^\prime R}\frac{\Gamma^a_\bk\Gamma^b_\bk}{\vert\bG_\bk\vert^2}
 \sigma^b_{\gamma^\prime\gamma}\biggr].
\end{displaymath}
Collecting the above expressions we have
\begin{displaymath}
 \langle\psi\vert\bS\bS_e\vert\psi\rangle=\frac{-3/4\sum_\bk\kappa^2_++1/4
 \sum_\bk\kappa^2_-}{\sum_\bk(\kappa^2_++\kappa^2_-)},
\end{displaymath}
and
\begin{align}
 \langle\psi\vert&(\bS+\bS_e)^2\vert\psi\rangle-\hFS\bS_e^2\FS=
 \frac{2\sum_\bk\kappa^2_-}{\sum_\bk(\kappa^2_++\kappa^2_-)}- \nonumber \\
 &-\frac{\sum_{\bk,\lambda}\theta(\ve_F-\ve_{\bk\lambda})
 (\kappa^2_++\kappa^2_--6\lambda\kappa_+\kappa_-)}
 {4\sum_\bk(\kappa^2_++\kappa^2_-)}, \nonumber
\end{align}
with the second term in the r.h.s. coming from
$\langle\psi\vert\bS^2_e\vert\psi\rangle$. In the absence of spin-orbit band
splitting $\kappa_-\equiv0$ and the above expression implies complete
screening. In the presence of SOI the change in the total spin is also finite
and for cases (i) and (ii) considered above
$\langle\psi\vert(\bS+\bS_e)^2\vert\psi\rangle-\hFS\bS_e^2\FS\sim\Delta_{SO}^2$.
In principle this change can be determined from local magnetic measurements,
but more precise methods than the one used here may be needed to determine
spatial dependence of the spin-spin correlations.

\paragraph*{Impurity spin susceptibility.--}
Finally, we consider the linear susceptibility of the system. Since our focus
is on the effect of SOI, we shall make a simplifying assumption that the system
is either cubic or tetragonal with magnetic field pointing along the $c$-axis,
and that the dominant effect of the field is on the impurity spin. In both
cases the Hamiltonian, Eq. \eqref{kondo_hamiltonian}, acquires a perturbation
$\delta H=-\mu BS_z$, where $\mu=g\mu_B$, $\mu_B$ is the Bohr magneton and $g$
is the appropriate Land\'e factor.

In order to account for $\delta H$ we need to change $E\to E+hM$ in Eqs.
\eqref{wavefunction} and \eqref{yosida_eq} with $h\equiv\mu B/2$ and $M=\pm1$.
A solution is sought in the form:
$X_{M\alpha}=x_sY^s_{M\alpha}+x_tY^t_{M\alpha}$ with $Y^{s}$ ($Y^{t}$) the
normalized singlet (triplet with zero total spin $z$-projection) basis states.
\begin{displaymath}
 \begin{pmatrix}
  1-3P & Q \\
  -3Q & 1+P
 \end{pmatrix}
 \begin{pmatrix}
  x_s \\ x_t
 \end{pmatrix}
 =0,
\end{displaymath}
where $P=(J_K/4N)\sum_{\bk,M}\kappa_+(E+hM)$ and
$Q=(J_K/4N)\sum_{\bk,M}M\kappa_+(E+hM)$. To lowest order in $\mu B/\delta E$,
the ground state energy becomes $E=\varepsilon_F-\delta E-\mu^2B^2/8\delta E$.
Therefore changes in the Kondo energy \eqref{kondo_energy} are
straightforwardly reflected in the spin susceptibility
\begin{displaymath}
 \chi=-\partial^2E/2\partial B^2=\mu^2/8\delta E.
\end{displaymath}

\paragraph*{Discussion.--}
Stimulated by the interest in non-centrosymmetric $f$-electron materials
\cite{hewson-1993,kimura-2012}, we investigated the influence of the lack of
inversion symmetry on interaction between conduction and localized electrons by
studying a single impurity Kondo model with a SOI in the conduction band. Using
a simple variational framework \cite{mahan-1990,yosida-1966} we presented
results for the ground-state properties of the system, valid for {\it any} form
of SOI and band structure of the host metal, even in cases when one cannot
reduce the problem to a single-channel Kondo Hamiltonian. It is the variational
nature of our approach, what allows us to deal with a multichannel model. In
particular, we demonstrated that: (1) the SOI can lead to an {\it exponential}
change of the Kondo temperature; (2) as the SOI explicitly breaks $SU(2)$
symmetry the Fermi sea does not completely screen the impurity spin, allowing
an extra magnetic degree of freedom in the Kondo phase.

Although a similar exponential enhancement of the Kondo temperature was
found in Ref. \onlinecite{zarea-2011}, we note that their result is
physically different from ours. The reason for this distinction is the fact
that in Ref. \onlinecite{zarea-2011} the authors started from an Anderson model
and used a Schrieffer-Wolff transformation \cite{schrieffer-1966}. Although
this is the usual way to ``freeze'' charge fluctuations at the
impurity, in the presence of SOI it can lead to unexpected results, such as the
Dzyaloshinky-Moriya coupling between impurity and conduction electrons spins,
which appears because of virtual transitions of localized electrons into the
conduction band where they accumulate a phase due to SOI. On the contrary we
started with a Kondo model that includes only spin fluctuations. Thus
modifications to the Kondo energy, Eq. \eqref{kondo_energy}, compared to its
value in a centrosymmetric material originates {\it purely} from SOI.

Our findings lead to an intriguing question regarding the influence of SOI on
the physics of the spin-$1/2$ Kondo lattice model. It is known
\cite{hewson-1993,bernhard-2011} that the heavy-fermion (Kondo screened) state
competes with magnetic phases. In the presence of a SOI impurity spins are not
completely screened \cite{VAji:2008} and may order, thus leading to a
coexistence of the heavy fermion state and magnetism. We leave investigation of
this problem for a future work.

We acknowledge support by DOE via Grant DE-FG02-08ER46492 (L. I. and I. V.),
by the NSF via Grants DMR-1105339 (I. V.) and DMR-0906655 (D. F. A.). This work
started during the ICAM Cargese School funded in part by I2CAM via NSF Grant
DMR-0844115.

\end{document}